# Revisiting $N_2$-$N_2$ collisional linewidth models for S-branch rotational Raman scattering

Mark Linne[a,1], Nils Torge Mecker[a], Christopher J. Kliewer[b], David Escofet-Martin[a], Brian Peterson[a]

[a]*University of Edinburgh, School of Engineering, Institute for Multiscale Thermofluids, The King's Buildings, Edinburgh, EH9 3FD, United Kingdom*
[b]*Combuston Research Facility, Sandia National Laboratories, Livermore, California, 94551, USA*

## Abstract

This paper presents an evaluation of two commonly accepted Raman linewidth models typically used to fit CARS model parameters to data; the Modified Exponential Gap (MEG) and the Energy Corrected Sudden (ECS) models. The adjustable parameters in each model have been fit to published experimental linewidths, and comparisons are made to the various publications that have already provided similar adjustable constants. The approaches presented in the literature are discussed as are differences with the findings presented here.

## 1. Introduction

Short-pulse coherent anti-Stokes Raman spectroscopy (CARS) has been proven as a temperature measurement at high pressures by several groups. Work has been reported on moderate pressure experiments at room temperature in the time domain, and up to 900 K in the frequency domain, using femtosecond CARS (fs-CARS) and hybrid femtosecond/picosecond (fs/ps) CARS [1, 2, 3, 4, 5]. Recently, we reported hybrid fs/ps rotational CARS (fs/ps HR-CARS) measurements on $N_2$ at pressures up to 70 atm and temperatures up to 1000 K [6]. During that work, we found that the inferred HR-CARS temperatures agreed better with thermocouple measurements when using the linewidths measured by Kliewer *et al.* [7], rather than the commonly used Modified Exponential Gap (MEG) linewidth model. When using the MEG model it was possible to achieve very low residuals in the spectral fits and yet the inferred temperature would disagree somewhat with the thermocouple. Here, therefore, both the MEG model and the Energy Corrected Sudden (ECS) model are explored further.

---

[1]Corresponding author, mark.linne@ed.ac.uk

There are many examples of similar work in the literature, all of it aimed at reliable and high-fidelity linewidth models for general use. In terms of measurements most closely related to this work (see Figure 1), Rahn and coworkers reported $N_2$ Q-branch linewidths measured by inverse Raman [8, 9]. That work is relevant because they used it to develop and introduce the MEG model. Short-pulse, S-branch rotational CARS linewidth data for $N_2$, extracted from time-domain measurements, have recently been supplied by Miller *et al.* [1], Kliewer *et al.* [7], and Meißner *et al.* [10]. Those results agree mostly with each other, except for some random variations in data that are not explained or discussed. More recently Haller and Varghese [11] have described spectral measurements of O- and S-branch rotational linewidths in $N_2$ using spontaneous Raman. Their data disagree with the data of Kliewer *et al.* by roughly 15%. Although the difference is not large, it falls outside the experimental uncertainties and the reasons for the differences are not currently clear.

Sitz and Farrow [12] published pump-probe measurements of state-to-state rate constants for $N_2$-$N_2$ collisions at 1 bar and room temperature, for even $J$ values from $J = 0$ to 14, and most of those data will be included in this study (the linewidth results at initial $J = 2$ and 14 were extreme outliers and so they are not included). Linewidths extracted from the Sitz and Farrow data are also represented in Figure 1.

The pressures and temperatures reported in the publications that are included in our evaluations are listed in Table 1.

Table 1: Conditions reported in the work against which model results are compared.

| **Citation** | **Pressure** (bar) | **Temperature** (K) |
|---|---|---|
| Kliewer *et al.* [7] | 1 | 294, 395, 495, 661, 868, 1116, 1466 |
| Miller *et al.* [1] | 1, 2.5, 5, 10, 15, 20 | 295 |
| Meißner *et al.* [10] | 1 | 295, 500, 870, 1000, 1200, 1400, 1650, 1750, 1900 |
| Rahn and Palmer [8] | 1 | 295, 500, 750, 1000, 1500 |
| Sitz and Farrow [12] | 1 | 298 |

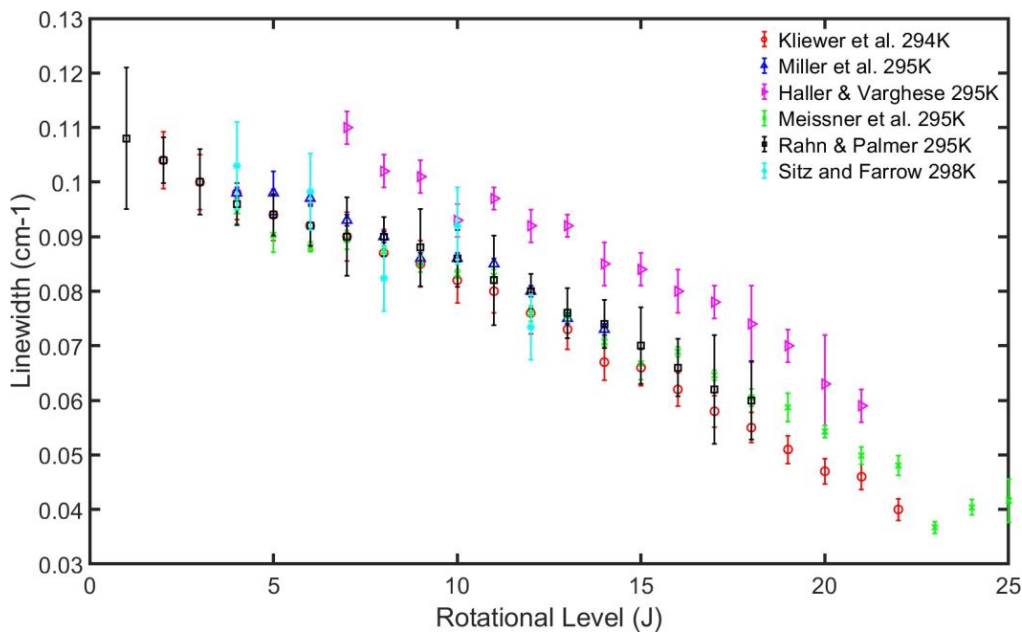

Figure 1: Experimental measurements of S-branch rotational Raman linewidths ($\Gamma$, FWHM), in $N_2$-$N_2$ mixtures, as a function of rotational quantum number ($J$) at 1 atm and 295 K by: ⓞ Kliewer *et al.* [7], ⏃ Miller *et al.* [1], ⊳ Haller and Varghese [11], x Meißner *et al.* [10], □ Rahn and Palmer [8] (taken from Q-branch measurements), and * selected data from Sitz and Farrow [12] (summations of state-to-state rates). The error bars are taken from the published uncertainties in each paper.

Since Rahn and co-workers [8, 9] introduced the MEG model in 1987, it has been evaluated against data by numerous authors (see, to name just a few, Lavorel *et al.* [13], Seeger *et al.* [14], Stauffer *et al.* [15], and the work cited above). In contrast, the Energy Corrected Sudden (ECS) collision model has many more variants and it is discussed in a large number of papers. As an example, the ECS-P law [16] assumes a power law dependence, the ECS-E law [17] assumes an exponential gap dependence, and ECS-EP [18, 19, 20] contains elements of both. Each of these versions has a number of variants.

Experimentalists fitting CARS spectra to extract temperature tend to use the linewidth approach that works best for them. In our prior work, for example, the measurements of Kliewer *et al.* [7] were used because they were the most closely related to the work performed by Mecker et al. [6], and by using them we were able to achieve better agreement with the thermocouple measurements. Others tend to use published models with published fitting parameters. Some perform their own fits. Still others have developed more complex libraries of models (sometimes selecting between MEG and ECS, depending on conditions) and fitting parameters. The specific model with specific fitting parameters that are known to work best at the experimental conditions are chosen in software. When used in this way, the models have become an informed interpolation between published experimental data points.

Competing concepts for linewidth computation would include modern machine learning techniques. They may provide good results at low computational cost. In addition, the simpler power law expansion proposed by Vispoel *et al.* [21] for microwave and infrared linewidths could potentially be adapted for Raman. Here we focus on MEG and ECS models, however, because they are the models used commonly in CARS codes.

## 2. Collisional line broadening and linewidth models

Because rotational Raman spectral lines are separated by as much as $\sim 8\,\mathrm{cm}^{-1}$ for nitrogen, it is generally accepted that line mixing can be neglected [22]. Haller and Varghese [11] have confirmed this assertion with measurements up to 70 atm. We therefore model isolated lines in what follows.

Isolated rotational Raman lines are broadened primarily via inelastic collision-induced changes in rotational states [17] and to a lesser extent to phase shifts arising from elastic reorientation of the molecular axis (arising from the anisotropic part of the Raman transition [23, 24]). There is a very small contribution of vibrational dephasing so it is neglected here [8]. Doppler broadening can contribute, especially at lower pressures and at high temperatures. Because our work has emphasized high pressures and moderate temperatures, Doppler broadening is not significant and it is also neglected in this work.

The inelastic collision-induced linewidth $\Gamma_j$ associated with a specific rotational level $j$ is typically written:

$$\Gamma_j = \sum_{i \neq j} \Gamma_{i,j} = \sum_{i \neq j} 2\gamma_{i,j} \tag{1}$$

where $i$ represents the other available rotational states, and $\Gamma_{i,j}$ represents a relaxation matrix containing the rates of collisional transfer from state $j$ into other states. Here $\Gamma_{i,j}$ represents the FWHM (full width at half-maximum) linewidth, while $\gamma_{i,j}$ represents the HWHM (half width at half-maximum) linewidth (e.g. the state-to-state rates measured by Sitz and Farrow).

Some papers apply the summation in equation 1 over all $J$ levels, without multiplying $\gamma_{i,j}$ by two. This approach can generate an error. For rotational Raman transitions, the selection rule is $\Delta J = \pm 2$. The sum in equation 1 is thus made only over the participating $J$ levels, not over all of them, and the $\times 2$ multiplier is included.

It is not possible within the combustion context to measure each of the $\Gamma_{i,j}$ using a line-integrated measurement of Raman spectra; the various contributors cannot be isolated from each other. State-to-state rates can be measured using techniques like double resonance spectroscopy (e.g. the pump-probe work by Sitz and Farrow [12]), but that approach requires exacting measurements (e.g. accurate timing of short pulses from multiple lasers) and it has not been pursued recently. Accurate and detailed trajectory calculations could be used, but they have also not been pursued. If one tried to develop scaling laws for each element of the relaxation matrix, the number of fitting parameters would make it intractable. It is therefore most common to use fitting laws having a small number of parameters to represent the large collection of these rates.

There are two general classes of fitting laws in use: 1) dynamic, angular momentum scaling models, and 2) statistical models. The various versions of the Energy Corrected Sudden (ECS) model represent the most common dynamic scaling laws and the Modified Exponential Gap (MEG) model is the most common statistical model. The MEG model is used most often because it is simpler and computationally faster. ECS models, on the other hand, include more of the underlying state-to-state physics and in principle it should require less ad-hoc manipulation. Both are evaluated here, mostly in the context of the measured linewidths by Kliewer *et al.* [7].

*2.1. MEG Model*

The MEG model assumes that the rates of collisional population transfer are based on the energy gap ($\Delta E_{ji}$) between the two states ($j$ and $i$). Some older models assume a simple power law in $\Delta E_{ji}$ but the MEG model assumes the dependence is exponential, similar to an Arrhenius-type equation in kinetic theory (note that $\gamma_{j,i}$ represents a rate, with units of bandwidth in $cm^{-1}$):

$$\gamma_{j,i} = \alpha P exp\left(-\frac{\beta \Delta E_{ji}}{kT}\right) \qquad (2)$$

where $a$ is an adjustable parameter (with typical units of $cm^{-1} atm^{-1}$), $P$ is pressure (usually in $atm$), $\beta$ is a unit-less adjustable parameter, $k$ is the Boltzmann constant, and $T$ is temperature. For various reasons this simple model failed to predict trends in measured linewidths, so additional scaling arguments based on physical reasoning were introduced (see Koszykowski *et al.* [9] and Rahn and Palmer [8]). The most common MEG model, written here for upward transitions ($i < j$), is:

$$\gamma_{j,i} = \alpha P \left(\frac{T_o}{T}\right)^n \left(\frac{1 + \frac{1.5 E_i}{kT\delta}}{1 + \frac{1.5 E_i}{kT}}\right)^2 exp\left(-\frac{\beta \Delta E_{ji}}{kT}\right) \qquad (3)$$

where $T_o$ is a reference temperature (usually 295 K), $n$ is a unit-less adjustable parameter, $E_i$ is the rotational energy of level $i$, and $\delta$ is another unit-less adjustable parameter. The temperature ratio taken to the $n$ power in equation 3 does a better job of representing the rates at higher temperature, while the more complex term containing rotational energy and $\delta$ helps match an experimentally observable change in the slope of the $\Gamma$ vs. $J$ curve at around $J = 6$. Sometimes equation 3 includes another temperature dependent factor:

$$f(T) = \frac{1 - e^{-m}}{1 - e^{-m(T/T_o)}} \qquad (4)$$

in which $m$ is another adjustable parameter (when equation 4 is used in equation 3, the value of $n$ is held fixed at 0.5). This expression was suggested by Farrow *et al.* [25] and it was intended to deal with fits at even higher temperatures. For downward transitions one uses detailed balance:

$$\gamma_{i,j} = \frac{2J_j + 1}{2J_i + 1} \gamma_{j,i} exp\left(-\frac{\Delta E_{ji}}{kT}\right) \qquad (5)$$

where $J_i$ is the rotational quantum number of level $i$. These equations are used to fill out the relaxation matrix, with diagonal elements set equal to zero in the absence of line mixing. The linewidths $\Gamma_j$ are then found using equation 1.

Martinsson *et al.* [26] apply the random phase approximation (RPA) [24] in an attempt to include the inelastic, anisotropic contributions to linewidth:

$$\Gamma = 0.5(\Gamma_J + \Gamma_{J+2}) \qquad (6)$$

Further evaluation is required to determine whether this term adequately corrects for the anisotropic contribution.

In the work reported here, the MEG model was implemented in Matlab and for the most part it was compared to the S-branch, time-domain fs/ps HR-CARS data of Kliewer *et al.* [7], which cover seven temperatures between 294 and 1466 $K$, spanning from $J = 2$ up to a maximum of $J = 42$ at the highest temperature, with low uncertainty ($\pm 5\%$). For the relaxation matrix developed here, the model contains rotational levels from $J = 0$ to $J = 70$ for all temperatures. It is simply necessary to ensure that all populated levels at the temperature of interest are accounted for, but going up to $J = 70$ has little cost in computing time and it guarantees that all necessary levels are included. A Levenberg–Marquardt nonlinear least squares fit was performed to the dataset of Kliewer *et al.* Fits were performed individually at each temperature, to extract best-fit values for the adjustable parameters in the MEG model. A first case utilized equations 1, 3, 5, and 6, while a second case utilized equations 1, 3, 4, 5, and 6. We found that using equation 4 generated better agreement across the range of temperatures, and so it was included.

The best-fit values of $\beta$ and $\delta$ did not vary much with temperature, and so they were averaged. The value of $m$ was then adjusted based upon temperature variations in the fitted values of $\alpha$ (which is supposed to be constant). Temperature variations are also represented in $\beta$ and $\delta$, however, and those are not directly fit. The values of the three ($m$, $\alpha$ and $\beta$) were then systematically adjusted to minimize overall disagreement with the entire dataset.

## 2.2. ECS-EP Model

The infinite order sudden (IOS) theory describes angular momentum coupling from one rotational state to another rotational state during a molecular collision. The effect of interaction potentials between the collision partners is neglected, and IOS assumes a diatomic molecule will not rotate during the collision. Scaling laws are therefore applied to the IOS model, to correct for such interactions and to accommodate finite collision durations [16]. The best known examples are the many forms of the energy corrected sudden (ECS) model. The concept underlying ECS scaling laws is an assumption that the entire relaxation matrix can be described by scaling most of the matrix using a known subset of it, usually one row (called the "base rate constants"). The base rates are typically not measured, so models have been developed to estimate them. The various versions of ECS differ based upon how the base rates are introduced and how the collision description is adjusted (using "adiabatic correction terms").

Millot [18] and Knopp *et al.* [20] describe a fairly common form of ECS which will be used here. For both upward and downward transitions (here $i$ denotes the rotational quantum number for the initial state and $j$ denotes the rotational quantum number for the final state, and $i \neq j$):

$$\gamma_{j,i} = (2j+1)\frac{\rho_{J_>}}{\rho_i}\sum_l (2l+1)\begin{pmatrix} i & l & j \\ 0 & 0 & 0 \end{pmatrix}^2 \frac{\Phi_l(\omega_{ij})^2}{\Phi_l(\omega_{l0})^2}Q_l \tag{7}$$

where $J_>$ denotes the larger of $i$ or $j$, and $\rho$ is the population density of the state given by Boltzmann statistics. The ECS model guarantees detailed balance by couching the population terms in this way. The term written as $\begin{pmatrix} i & l & j \\ 0 & 0 & 0 \end{pmatrix}$ is a Wigner 3-$j$ symbol describing coupling between various rotational levels. Here $l$ is a dummy representation for the rotational levels over which the sum is carried. The lower row in the 3-$j$ symbol contains the associated magnetic quantum numbers. Here they are set to zero based on the trajectory used in IOS [17]. The upper row in the 3-$j$ symbol is sometimes written with the rotational quantum numbers in a different order, but when the lower row is all zeroes the terms with different ordering are equal [27]. The term $\Phi_l(\omega_{ij})$ is an adiabatic correction given by:

$$\Phi_l(\omega_{ij}) = \left(1 + (\omega_{J_>} - \omega_{J_{>-\Delta}})^2 \frac{\tau^2}{24}\right)^{-1} \tag{8}$$

where $(\omega_{J_>} - \omega_{J_{>-\Delta}})$ represents an energy gap (e.g. $\omega = E/\hbar$) to the nearest transition where $\Delta = 2$ for homo- and $\Delta = 1$ for hetero-nuclear molecules. Recall that $J_>$ is the larger of $i$ or $j$, so equation 8 can also be used for $\Phi_l(\omega_{l0})$. The collision time is expressed by:

$$\tau = \frac{l_c}{\bar{v}} \tag{9}$$

where $l_c$ is a representative interaction length (adjustable, with typical units of $m$), and $\bar{v}$ is the average molecular velocity ($m/s$) given by the Maxwellian distribution. Finally, $Q_f$ represents a model for the base rates for $\pounds \rightarrow 0$ here:

$$Q_{l \rightarrow 0} = -\alpha \left(\frac{T_o}{T}\right)^n [l(l+1)]^{-\gamma} exp\left(-\beta \frac{E_l}{kT}\right) \qquad (10)$$

where $[l(l+1)]^{-\gamma}$ limits angular momentum transfer for small energy gaps (the "power-gap law", ECS-P), and the exponential term scales the rates of transfer based on the energy gap (ECS-E). The combination of the two, as represented in equation 10, is termed the ECS-EP law. ECS-EP has five adjustable constants $l_c$, $a$, $n$, $\gamma$ and $\beta$. The term $a$ is an adjustable parameter (with typical units of $cm^{-1} atm^{-1}$) and so the equations are written for atmospheric pressure. As with MEG, $\gamma_{j,i}$ will scale with pressure.

Here the ECS-EP model (equations 7 to 10) was implemented in Matlab and for the most part it was compared to the same S-branch, time domain fs/ps HR-CARS data of Kliewer *et al.* [7], similar to the MEG section. As before, the model contains rotational levels from $J = 0$ to $J = 70$ for all temperatures. To implement equation 7, a Matlab script named 'Wigner3j' was validated against known results and then implemented in the ECS-EP script. A Levenberg-Marquardt nonlinear least squares fit was performed to match the dataset of Kliewer *et al.* Millot [18] points out, however, that "there are correlations between the four parameters of the ECS-EP law" and he argues that one must limit some of the parameters while fitting the others. We did indeed encounter this issue, so only one adjustable constant was varied at a time, while the others were held fixed. The values of the five constants were then consecutively adjusted in a full model to minimize overall disagreement with the entire dataset. Unfortunately, such an approach cannot be used to ensure that the global optimum has been reached. There is a possibility that our results represent a sub-optimum.

Other issues are associated with this model. We have found that different algorithm structures can generate differences in linewidths on the order of 10% to 20%. Here, therefore, we explain in detail how our code was structured. In equation 7, early authors perform the summation over l by summing over $|j - i|$ to $j + i$ (with justifications), while others sum over all levels included in the model (e.g. over 0→70 here). We have evaluated both approaches, but we get better results using publicaed ECS-EP constants if we use 0→70. Moreover, equation 10 does not allow the use of $l = 0$ because the term goes to infinity. For this implementation, the first entry in the summation is missing the $[l(l+1)]^{-\gamma}$ term (e.g. the ECS-E model is used for that one entry), but it appears in subsequent entries of the summation. In addition, the term ($\omega_{J_>} - \omega_{J_> - \Delta}$) in equation 8 also has a problem at low $J$. For nitrogen the nearest transition is for $\Delta = 2$. When $J$ is between 0 and 2, that difference term can't be used (because we do not have access to negative rotational quantum numbers). For this reason, we do not include that term until $J = 3$. These issues mean that the calculation is different for $J$ (or $l$) = 0 up to $J$ (or $l$) = 2. The slope of computed linewidths vs. $J$ at low $J$ is somewhat different, therefore, from the slope after $J = 2$, but low $J$ terms must be included in order to fill the matrix. Finally, we apply the RPA (equation 6) to our estimates.

## 3. Results

### 3.1. Fits to the MEG Model

Example values for adjustable constants reported in the literature, together with the fit values produced here, are provided in Table 2. Lavorel *et al.* provide four cases but only two are shown here for reference. Note that Koszykowski *et al.* and Lavorel *et al.* were fitting rotational linewidths in Q-branch vibrational spectra, while Martinsson *et al.* and Stauffer *et al.* were fitting S-branch purely rotational linewidths. Most of the sources also include uncertainties, although at times their uncertainties are smaller than the differences between publications. The authors in rows one through four do not provide an explanation of how they estimated uncertainties. Our uncertainties are based on standard deviations across the seven temperatures used by Kliewer et al. [7], without accounting for changes made during systematic adjustment of $m$, $a$ and $\beta$. Our reported uncertainties are therefore approximate.

The last two columns of Table 2 contain a comparison between model results and experiments when using the constants in each row. The numbers under the column for $\Delta_K$ represent average percent disagreement with the entire dataset published by Kliewer *et al.* while the $\Delta_F$ column contains average disagreement with the state-to-state rates published by Sitz and Farrow [12] (the data at $J = 2$ and $J = 14$ were not included). Note that for the models of Koszykowski *et al.*, Martinsson *et al.*, and Lavorel I *et al.* we used $f(T) = 1$ and for the model of Stauffer *et al.* and Lavorel II we used equation 4 with their suggested value for $m$.

Use of the new constants generated in this work (shown in the bottom row of Table 2) generates an average disagreement with the data that falls under the experimental uncertainty of the data published by Kliewer et al. We also find good agreement with the state-to-state rates published by Sitz and Farrow (we fall within their uncertainties), validating the code itself. Figure 2 contains a plot of the data of Kliewer et al. against these new MEG results, and one can see that the model works well. The greatest disagreement with data is at 294 *K*. This low temperature disagreement is a known issue and it is one of the reasons others use bespoke models for each set of conditions.

Table 2: MEG model constants
($\Delta_K$ indicates average disagreement between the model and the entire linewidth dataset from Kliewer et al. [7] while $\Delta_F$ indicates average disagreement between the model and the relaxation matrix elements of Sitz and Farrow [12])

| **Citation** | $a$ | $\beta$ | $\delta$ | $n$ | $m$ | $\Delta_K$ (%) | $\Delta_F$ (%) |
|---|---|---|---|---|---|---|---|
| Koszykowski [9] | 0.023 ± 0.003 | 1.67 ±0.15 | 1.26 ± 0.06 | 1.346 ± 0.006 | 0 | 7.8 | 3.3 |
| Martinsson [26] | 0.02645 ± 0.00026 | 1.890 ±0.018 | 1.174 ± 0.029 | 1.365 ± 0.005 | 0 | 5.7 | 5.3 |
| Lavorel I [13] | 0.02648 ± 0.00026 | 1.894 ±0.029 | 1.175 ± 0.017 | 1.366 ± 0.005 | 0 | 5.5 | 5.3 |
| Lavorel III [13] | 0.02646 ± 0.00021 | 1.850 ±0.024 | 1.199 ± 0.015 | 0.5 | 0.1381 ± 0.0031 | 6.3 | 7.8 |
| Stauffer [15] | 0.023 | 1.67 | 1.21 | 0.5 | 0.1487 | 9.5 | 5.3 |
| This work | 0.0283 ± 0.002 | 1.942 ±0.03 | 1.259 ± 0.07 | 0.5 | 0.1360 ± 0.0001 | 3.3 | 4.5 |

The level of disagreement between the various models and the data contained in Table 2 is low. As Afzelius *et al.* [24] point out, however, there will be roughly 10% disagreement at low $J$ levels caused by the fact that the MEG model assumes that broadening arises entirely from inelastic collisional population transfers (appropriate for Q-branch transitions in vibrational CARS), and the RPA is used in an attempt to modify it for use on purely rotational transitions. In the process, re-orientations of molecular axes are not adequately accounted for even though they contribute to rotational linewidths, especially at low $J$. One can see this effect in Figure 2, especially at 294 *K*.

This new MEG model was also compared to the S-branch, time-domain fs/ps HR-CARS data of Miller *et al.* [1], who published 295 *K* data over six pressures (1, 2.5, 5, 10, 15, and 20 bar). Comparison across all of the pressures reported by Miller *et al.* generated an average disagreement with their data of 5.0%. Afzelius *et al.* [24] mention that MEG model errors increase with pressure. There was a slight trend upwards in disagreement between the model and the data as pressure increased, but it was not strong and the disagreements remained close to the published uncertainties of Miller *et al.* Comparing this MEG model to the data of Meißner *et al.* [10] generates an average difference of 6.7%. Random variations in the Meißner data are high, sometimes falling outside of their uncertainty bars, and so one would expect higher disagreement with a smooth, modeled curve. Comparisons were also made to the Q-branch data of Rahn and Palmer [8] where this model produced 6.1% disagreement.

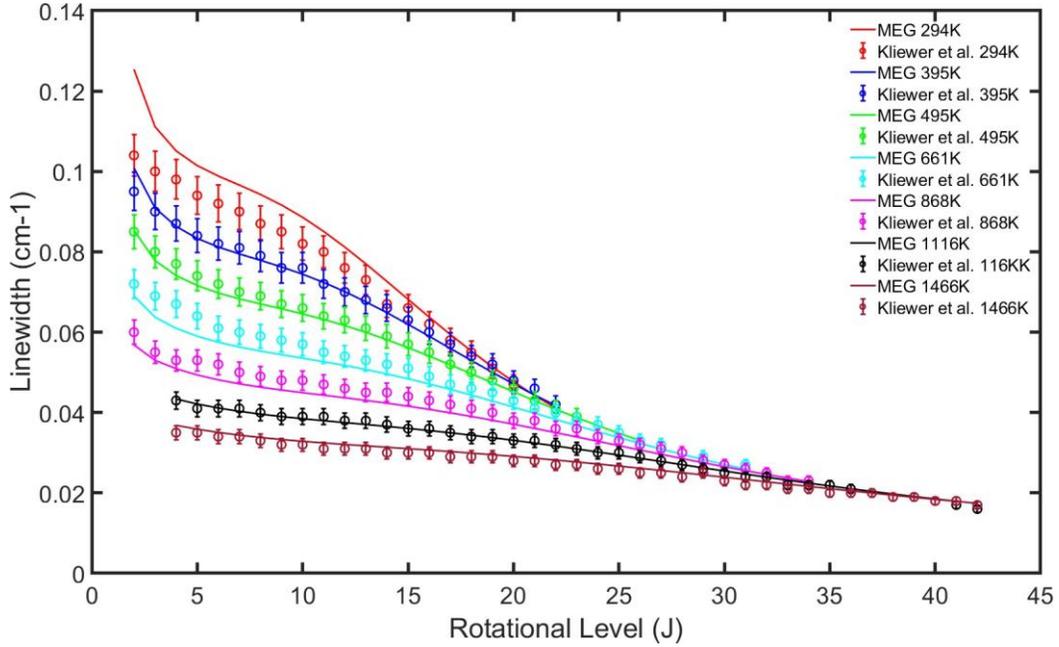

Figure 2: MEG model fits to the data of Kliewer *et al.* at various temperatures, using the constants in the bottom row of Table 2.

*3.2. Fits to the ECS-EP Model*

Example values for adjustable constants reported by Millot [18], together with the fit values produced here, are provided in Table 3. Only one published set of constants is included because there are many variations of the ECS model with somewhat different details, and we have chosen to investigate only one formalism. Furthermore, because the low $J$ levels in this model are treated differently, we do not include them in the comparison to data and they were not used to set the final constants for this work, because they skew the fits if they are included. For this reason the quoted disagreement is low, but that happens because we are comparing a more limited set of data. Note also that we do not show comparison to the individual relaxation matrix elements of Sitz and Farrow [12] because the low $J$ terms are not included in evaluations (see discussion just below). Note that Millot [18] fitted rotational linewidths in Q-branch vibrational spectra.

Table 3: ECS-EP model constants
($\Delta_K$ indicates average disagreement between the model and the entire linewidth dataset from Kliewer et al. [7])

| Citation | $a\,(cm^{-1}atm^{-1})$ | $\beta$ | $\gamma$ | $l_c\,(m)$ | $n$ | $\Delta_K$ (%) |
|---|---|---|---|---|---|---|
| Millot [18] | 0.0189 ± 0.0006 | 0.1309 ±0.0062 | 0.7419 ± 0.0121 | 3.5×10⁻¹⁰ ± 0.037×10⁻¹⁰ | 1.112 ± 0.015 | 3.8 |
| This work | 0.0561 ± 0.0005 | 1.89 ±0.05 | 0.939 ± 0.009 | 0.98 × 10⁻¹⁰ ± 0.05×10⁻¹⁰ | 1.117 ± 0.005 | 2.3 |

Use of the new constants generated in this work (shown in the bottom row of Table 3) generates an average disagreement with the more limited range of data that falls under the experimental uncertainty of the data published by Kliewer et al. We find that agreement with the state-to-state rates published by Sitz and Farrow is not good (61% disagreement). The same is true for the constants published by Millot (62% disagreement). This problem may be caused by treatment of the low $J$ levels, but they must be dealt with in a special way no matter which algorithm is used. Alternatively, the point made by Millot that the constants are coupled might lead to sub-optimum fits that fail to reproduce the rates. Both sets of constants give good agreement with the measured linewidths (over the more limited range) while disagreeing with the measured state-to-state rates. Note also that there is a significant discrepancy between our constants and those of Millot, likely due, again, to the coupling of constants. Figure 3 contains a plot of the data of Kliewer et al. against these new ECS-EP results, and one can see that the model works well at high $J$ levels but it fails at 294 $K$, and from 495 $K$ to 1116 $K$ the model falls outside the data for $J \leq 5$.

This new ECS model was also compared to the same S-branch, time-domain fs/ps HR-CARS data of Miller *et al.* [1] covering six pressures (1, 2.5, 5, 10, 15, and 20 bar). Comparison across all of the pressures reported by Miller *et al.* generated an average disagreement with their data of 7.4%.

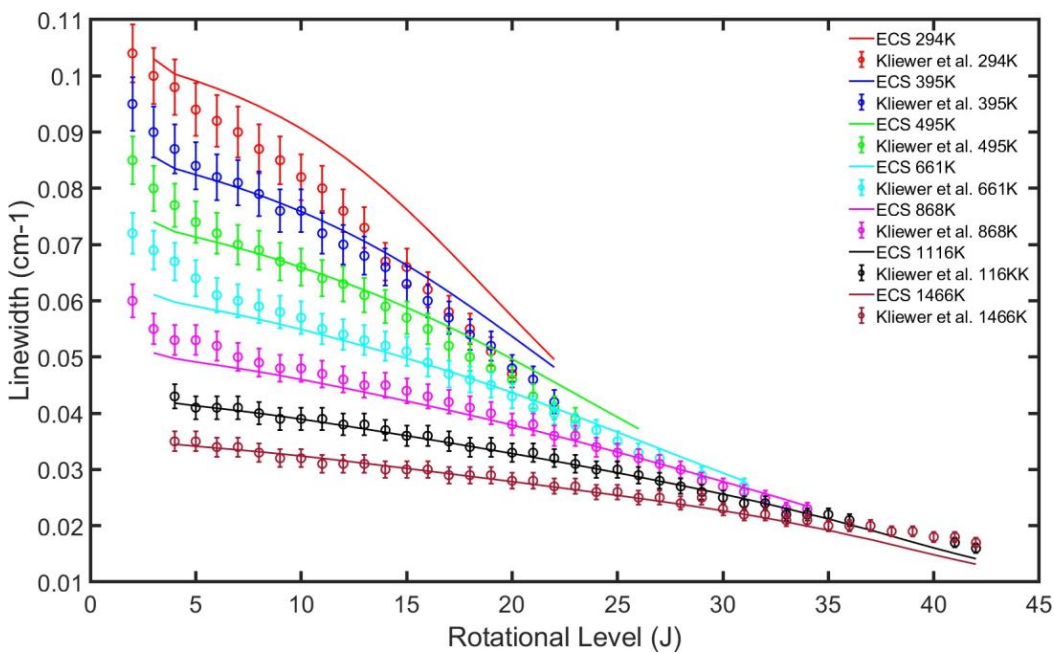

Figure 3: ECS-EP model fits to the data of Kliewer *et al.* at various temperatures, using the constants in the bottom row of Table 3.

Comparing this ECS model to the data of Meißner *et al.* [10] generates an average difference of 5.5%. Comparisons were also made to the Q-branch data of Rahn and Palmer [8], where this model produced 6.7% disagreement. Note again that the low $J$ values have not been included here so the disagreement values are lower than they would have been if the low $J$ values had been included.

## 4. Conclusions

More recent time-domain measurements of pure rotational S-branch linewidths have been evaluated in this work using two more common models and good agreement has been achieved. These results have been based on a nitrogen collisional environment. With respect to combustion measurements, the assumption of simplified collisional partners for thermometry in fs/ps HR-CARS, over a range of collisional partners and probe delays, has been applied successfully by Bohlin *et al.* [28, 29] and by Retter *et al.* [30].

ECS theories have been developed with the goal to provide a more predictive model for linewidths. While the goal is admirable, a number of adaptations with adjustable constants have been required, and one ends up fitting the same number of constants as one fits for the MEG model. Unfortunately, there is some cross-talk between constants in the ECS-EP model and so fits are not as straightforward as one would like. There is also an issue of how to deal with very low $J$ equations, where correct treatment matters the most. In addition, while good agreement with linewidths can be achieved, the ECS-EP model implemented here did not reproduce measured state-to-state rates very well. Finally, the ECS models run more slowly (almost 10 times more slowly) than the MEG model. This happens in part because the Wigner 3-$j$ values must be calculated repeatedly for every summation, and there are many summations.

The MEG model generated in this work was straightforward to implement, provided good agreement with measurements (including state-to-state rates), and it is relatively fast. Speed becomes important when, for example, one is analyzing numerous measurements using fs/ps HR-CARS in a line-format, to extract temperature vs. position and temperature gradients.

In answer to the question regarding low residuals in spectral fits, but with poor temperature agreement; we are of the opinion that sufficiently many parameters are fit when modelling fs/ps HR-CARS spectra that the code can overcome small problems with linewidth models to produce low residuals, but the inferred temperature can be less accurate. Our guidance would be to use the MEG model with the constants provided herein when performing pure rotational CARS in combustion experiments, especially fs/ps HR-CARS. We make that suggestion based upon the fact that good temperature agreement with the measured linewidths of Kliewer *et al.* [7] has been achieved [6]. These new MEG model constants reproduce the data of Kliewer *et al.* within the experimental uncertainty in almost all cases, and so it provides a good computational representation of the dataset. Larger deviations were observed for the coldest data, and future research should be devoted to understanding why the scaling models begin to deviate at low temperatures. Vibrational Raman would likely require different fits, and for atmospheric pressure or above one needs to invoke the G-equation formalism to include line mixing in that case.

The ECS theories represent an attempt to use classical trajectory theory to predict linewidths, based upon first principles (see e.g. the work of Bonamy and co-workers [31, 32, 33, 23, 13]). Researchers who are deeply involved in that area can develop specialized adjustments that improve performance, but for combustion diagnostics this is not a convenient option.

For the future, it is probably best to bifurcate. Those needing to use a combustion diagnostic in the near term can do well using the MEG model. To develop more predictive theories, modern computing offers an opportunity to generate much more accurate trajectories, and potentially to use those results to inform a model like ECS. Alternatively, experimental data or trajectory calculations could be used with machine learning techniques, or to find simpler power law expansions [21].

## 5. Acknowledgements

B. Peterson and D. Escofet-Martin gratefully acknowledge financial support from the European Research Council (ERC grant number 759546).

This material is based upon work supported by the Division of Chemical Sciences, Geosciences and Biosciences, Office of Basic Energy Sciences (BES), USDOE. Sandia National Laboratories is a multimission laboratory managed and operated by National Technology and Engineering Solutions of Sandia, LLC, a wholly owned subsidiary of Honeywell International, Inc., for the USDOE National Nuclear Security Administration under Contract No. DE-NA0003525. This paper describes objective technical results and analysis. Any subjective views or opinions that might be expressed in the paper do not necessarily represent the views of the USDOE or the U.S. Government.

# References

[1] J. D. Miller, S. Roy, J. R. Gord, T. R. Meyer, Communication: Time-domain measurement of high-pressure $N_2$ and $O_2$ self-broadened linewidths using hybrid femtosecond/picosecond coherent anti-Stokes Raman scattering, J. Chem. Phys. 135 (20) (2011) 201104.

[2] P. J. Wrzesinski, H. U. Stauffer, W. D. Kulatilaka, J. R. Gord, S. Roy, Time-resolved femtosecond CARS from 10 to 50 Bar: collisional sensitivity, J. Raman Spectrosc. 44 (10) (2013) 1344–1348.

[3] G. Matthäus, S. Demmler, M. Lebugle, F. Küster, J. Limpert, A. Tünnermann, S. Nolte, R. Ackermann, Ultra-broadband two beam CARS using femtosecond laser pulses, Vib. Spectrosc. 85 (2016) 128–133.

[4] M. Kerstan, I. Makos, S. Nolte, A. Tünnermann, R. Ackermann, Two-beam femtosecond coherent anti-Stokes Raman scattering for thermometry on $CO_2$, Appl. Phys. Lett. 110 (2) (2017) 021116.

[5] S. P. Kearney, P. M. Danehy, Pressure measurements using hybrid femtosecond/picosecond rotational coherent anti-Stokes Raman scattering, Opt. Lett. 40 (17) (2015) 4082–4085.

[6] N. T. Mecker, T. L. Courtney, B. D. Patterson, D. Escofet-Martin, B. Peterson, M. Linne, C. J. Kliewer, Hybrid rotational femtosecond/picosecond coherent anti-Stokes Raman spectroscopy (HR-CARS) of nitrogen at high pressures (1-70 atm) and temperatures (300-1000 K), J. Opt. Soc. Am. B 37 (4) (2020) 1035–1046.

[7] C. J. Kliewer, A. Bohlin, E. Nordström, B. D. Patterson, P.-E. Bengtsson, T. B. Settersten, Time-domain measurements of S-branch $N_2$-$N_2$ Raman linewidths using picosecond pure rotational coherent anti-Stokes Raman spectroscopy, Appl. Phys. B 108 (2) (2012) 419–426.

[8] L. A. Rahn, R. E. Palmer, Studies of nitrogen self-broadening at high temperature with inverse Raman spectroscopy, J. Opt. Soc. Am. B 3 (9) (1986) 1164–1169.

[9] M. L. Koszykowski, L. A. Rahn, R. E. Palmer, Theortical and Experimental Studies of High-Resolution Inverse Raman Sepectra of $N_2$ at 1-10 atm, J. Phys. Chem. 91 (1987) 41–46.

[10] C. Meißner, J. I. Hölzer, T. Seeger, Determination of $N_2$-$N_2$ and $N_2$-$O_2$ S-branch Raman linewidths using time-resolved picosecond pure rotational coherent anti-Stokes Raman scattering , Applied Optics 58 (10) (2019) 47–53.

[11] T. W. Haller, P. L. Varghese, Measurements of pressure broadening of $N_2$ in the anisotropic tensor component of spontaneous Raman spectra , Combustion and Flame 224 (2021) 166–176.

[12] G. O. Sitz, R. L. Farrow, Pump–probe measurements of state-to-state rotational energy transfer rates in N2 (v=1), J. Chem. Phys. 93 (1990) 7883.

[13] B. Lavorel, L. Guillot, J. Bonamy, D. Robert, Collisional Raman linewidths of nitrogen at high temperature (1700 - 2400K), Optics Letters 20 (10) (1995) 1189–1191.

[14] T. Seeger, F. Beyrau, A. Bräuer, A. Leipertz, High-pressure pure rotational CARS: Comparison of temperature measurements with $O_2$, $N_2$ and synthetic air, J. Raman Spectrosc. 34 (12) (2003) 932–939.

[15] H. U. Stauffer, J. D. Miller, M. N. Slipchenko, T. R. Meyer, B. D. Prince, S. Roy, J. R. Gord, Time-and frequency-dependent model of time-resolved coherent anti-Stokes Raman scattering (CARS) with a picosecond-duration probe pulse, J. Chem. Phys. 140 (2) (2014) 024316.

[16] A. E. DePristo, S. D. Augustin, R. Ramaswarmy, H. Rabitz, Quantum number and energy scaling for nonreactive collisions, J. Chem. Phys. 71 (2) (1979) 850–865.

[17] T. A. Brunner, D. Pritchard, Fitting laws for rotationally inelastic collisions, in: P. K. Lawley (Ed.), Dynamics of the Excited State, John Wiley and Sons, Hoboken, NJ, 1982.

[18] G. Millot, Rotationally inelastic rates over a wide temperature range based on an energy corrected sudden-exponential-power theoretical analysis of Raman line broadening coefficients and Q branch collapse, J. Chem. Phys. 93 (11) (1990) 8001–8010.

[19] G. Millot, R. Saint-Loup, J. Santos, R. Chaux, H. Berger, Collisional effects in the stiulated Raman Q branch of $O_2$ and $O_2$-$N_2$, J. Chem. Phys. 96 (2) (1992) 961–971.

[20] G. Knopp, P. Radi, M. Tulej, T. Gerber, P. Beaud, Collision induced rotational energy transfer probed by time-resolved coherent anti-Stokes Raman scattering, J. Chem. Phys. 118 (18) (2003) 8223–8233.

[21] B. Vispoel, J. H. Cavalcanti, E. T. Paige, R. R. Gamache, Vibrational dependence, temperature dependence, and prediction of line shape parameters for the $H_2O$–$N_2$ collision system, Journal of Quantitative Spectroscopy and Radiative Transfer 253 (2020) 107030.

[22] J. Bood, P.-E. Bengtsson, T. Dreier, Rotational coherent anti-Stokes Raman spectroscopy (CARS) in nitrogen at high pressures (0.1-44 MPa): experimental and modelling results, J. Raman Spectrosc. 31 (8-9) (2000) 703–710.

[23] L. Bonamy, J. Bonamy, D. Robert, S. Temkin, G. Millot, B. Lavorel, Line coupling in anisotropic Raman branches, J. Chem. Phys. 101 (9) (1994) 7350–7356.

[24] M. Afzelius, P.-E. Bengtsson, J. Bood, J. Bonamy, F. Chaussard, H. Berger, T. Dreier, Dual-broadband rotational CARS modelling of nitrogen at pressures up to 9 MPa. II Rotational Raman line widths, Appl. Phys. B 75 (2002) 771–778.

[25] R. L. Farrow, R. Trebino, R. E. Palmer, Measurements of pressure broadening of $N_2$ in the anisotropic tensor component of spontaneous Raman spectra , Appl. Opt. 26 (2021) 166–176.

[26] L. Martinsson, P.-E. Bengtsson, M. Aldén, S. Kröll, A test of different rotational Raman linewidth models: Accuracy of rotational coherent anti-Stokes Raman scattering thermometry in nitrogen from 295 to 1850 K, J. Chem. Phys. 99 (4) (1993) 2466–2476.

[27] R. N. Zare, Angular Momentum: Understanding Spatial Aspects in Chemistry and Physics, John Wiley & Sons, New York, 1988.

[28] A. Bohlin, M. Mann, B. D. Patterson, A. Dreizler, C. J. Kliewer, Development of two-beam femtosecond/picosecond one-dimensional rotational coherent anti-stokes raman spectroscopy: Time-resolved probing of flame wall interactions, Proceedings of the Combustion Institute 35 (2015) 3723–3730.

[29] A. Bohlin, C. Jainski, B. D. Patterson, A. Dreizler, C. J. Kliewer, Multiparameter spatio-thermochemical probing of flame–wall interactions advanced with coherent raman imaging, Proceedings of the Combustion Institute 36 (2017) 4557–4564.

[30] J. E. Retter, G. S. Elliott, S. P. Kearney, Dielectric-barrier-discharge plasma-assisted hydrogen diffusion flame. Part 1: Temperature, oxygen, and fuel measurements by one-dimensional fs/ps rotational CARS imaging, Combustion and Flame 191 (2018) 527–540.

[31] D. Robert, J. Bonamy, Short range firce effects in semiclassical molecular line broadening calculations, Le Journal De Physique 40 (10) (1979) 923 – 943.

[32] B. Lavorel, G. Millot, J. Bonamy, D. Robert, Study of Rotational Relaxation Fittign Laws from Calculation of SRS $N_2$ Q-Branch, Chem. Phys. 115 (1987) 69 – 78.

[33] L. Bonamy, J. Bonamy, D. Robert, B. Lavorel, R. Saint-Loup, R. Chaux, J. Santos, H. Berger, Rotationally inelastic rates for $N_2$-$N_2$ system from a scaling theoretical analyis of the stimulated Raman Q branch, J. Chem. Phys. 89 (1988) 5568 – 5577.